\documentclass[superscriptaddress,onecolumn,preprint ]{revtex4-1}
\usepackage{graphicx}
\usepackage{amssymb}
\usepackage{epstopdf}
\usepackage{amsmath,dsfont, commath}
\usepackage{bm}
\usepackage{braket}
\usepackage{changes}
\usepackage[capitalize]{cleveref}% load last

\newcommand{\be}{\begin{equation}}
	\newcommand{\ee}{\end{equation}}
\newcommand{\bea}{\begin{eqnarray}}
	\newcommand{\eea}{\end{eqnarray}}
\newcommand{\ba}{\begin{array}}
	\newcommand{\ea}{\end{array}}

\newcommand{\se}{Schr\"{o}dinger equation}
\newcommand{\bl}{\begin{flalign}}
	\newcommand{\enl}{\end{flalign}}

\newcommand{\mc}[1]{\mathcal{#1}}

\newcommand{\tdse}{time-dependent Schr\"{o}dinger equation}

\newcommand{\eq}[1]{Eq. \eqref{#1}}

\newcommand{\Eq}[1]{Equation \eqref{#1}}
\newcommand{\fig}[1]{Fig. (\ref{#1})}
\renewcommand{\sec}[1]{Sec. \ref{#1}}

\newcommand{\half}{\frac{1}{2}}

\newcommand{\proj}[1]{\ket{#1}\bra{#1}}

\renewcommand{\bf}[1]{\mathbf{#1}}
\newcommand{\grad}{\nabla}

\begin{document}

	%opening
	\title{Nonadiabatic conical intersection  dynamics in the local diabatic representation with Strang splitting and Fourier basis }%coupled electron-nuclear motion}

\author{Bing Gu}
\email{gubing@westlake.edu.cn}
\affiliation{Department of Chemistry \& Department of Physics, Westlake University, Hangzhou, Zhejiang 310030, China}
\affiliation{Institute of Natural Sciences, Westlake Institute for Advanced Study, Hangzhou, Zhejiang 310024, China}
\begin{abstract}

	We develop and implement an exact conical intersection nonadiabatic wave packet dynamics method that combines the local diabatic representation, Strang splitting for the total molecular propagator, and   discrete variable representation with uniform grids. By employing the local diabatic representation, this method captures all non-adiabatic effects including nonadiabatic transitions, electronic coherences, and geometric phase. Moreover, it is free of singularities in the first and second derivative couplings, and does not require a smooth gauge of electronic wavefunction phase.
	  %This alleviates signnifcantly the exponential scaling of computation scost with respect oteh size of moelcules,
	  We further show that in contrast to the adiabatic representation, the split-operator method can be directly applied to the full molecular propagator  with the locally diabatic ansatz.
	  The Fourier series, employed as the primitive nuclear basis functions, is universal and can be applied to all types of reactive coordinates.
	  The combination of local diabatic representation, Strang splitting, and Fourier basis allows exact modeling of conical intersection quantum dynamics directly with adiabatic electronic states that can be obtained from standard electronic structure computations.

\end{abstract}
\maketitle

\section{Introduction}

Conical intersections (CIs), a cone-like structure formed in molecular configuration space where adiabatic potential energy surfaces of electronic states intersect,  play a critical role in understanding and explaining virtually all photochemical and photophysical phenomena such as vision, photostability of DNA and RNA nucleobases, Jahn-Teller distortion, photoisomerization, light-driven dynamics, singlet fission, ultrafast spectroscopy,  polaritonic chemistry \cite{mead1982, gu2023, domcke2011, casida2012, polli2010, mandal2023, gu2020c, rafiq2023, quenneville2003, gu2021b, casida2012, chen2022e}.
%At a conical intersection, two or more electronic states come together in such a way that their potential energy surfaces touch at a single point, forming a cone-like structure. 
In the vicinity of CIs,  electronic and nuclear motion become strongly correlated as their energy scales becomes comparable, and hence the Born-Oppenheimer approximation which separates the electronic and nuclear motion breaks down.
% This distinctive geometry enables efficient interconversion between electronic states (i.e. nonadiabatic transitions), a process critical for various photochemical and biological reactions including vision, photodamage of DNA and RNA nucleobases.
 One intriguing aspect of CIs lies in their capacity to facilitate ultrafast transitions between electronic states, often occurring on femtosecond timescales.  
 %This nonradiative relaxation is a process critical for various photochemical, photophysical processes and  biological reactions including vision, photodamage of DNA and RNA nucleobases \cite{}.

%Understanding conical intersections has become imperative in unraveling the intricacies of molecular behavior, particularly in the context of excited-state dynamics. These intersections serve as gateways for electronic energy transfer, photoisomerization, and other photophysical processes that govern the behavior of molecules in excited states. Consequently, researchers delve into the exploration of conical intersections to decipher the mechanisms driving light-induced reactions and to design molecular systems with tailored photophysical properties.

%Understanding molecular motion is essential to understanding a large variety of problems in chemical physics, materials science.

Besides facilitating electronic transitions, conical intersections introduce a geometric complexity to the nuclear quantum dynamics \cite{ryabinkin2013, ryabinkin2017, mead1992}. The electronic wave functions acquires a geometric phase of $\pi$ as the nuclear coordinates traverse a closed path around a conical intersection, that is distinct from the dynamic phase associated with the time evolution of the wave function.
The geometric phase in conical intersections arises from the nontrivial geometry of the configuration space.  This phase is invariant to the deformation of the closed path taken in the nuclear space as long as it encircles the CI, reflecting the topological nature of the quantum evolution.
Understanding the role of geometric phase in conical intersections has profound implications for elucidating the mechanisms governing nonadiabatic transitions and excited-state dynamics and even for low-energy ground state dynamics where the CIs are energetically inaccessible \cite{xie2019}.
%It provides valuable insights into the geometric aspects of quantum wave function evolution and offers a unique perspective on the interplay between electronic and nuclear motion in molecular systems.

%In this adiabatic representation, the  ansatz for the molecular wavefunction is $\sum_n \ket{\phi_n} \chi_n(\bf R, t)$ where $\phi_n(\bf r)$ is the $n$th adiabatic electronic state. While this approach is useful for diabatic model systems, in the adiabatic representation, it is hindered by the divergences of derivative couplings at CIs.

%
Conventionally, the nonadiabatic conical intersection dynamics is either understood in the adiabatic or the diabatic representation \cite{yarkony2019}. In the adiabatic representation, the ansatz for  molecular wavefunction is expanded using the adiabatic electronic states and associated nuclear wavepackets, so-called Born-Huang expansion.  %$\sum_n \ket{\phi_n} \chi_n(\bf R, t)$ where $\phi_n(\bf r)$ is the $n$th adiabatic electronic state. 
This leads to an intuitive picture of nuclear wavepackets evolving on the adiabatic potential energy surfaces (APES), and makes electronic transitions only when it reaches a region with significant nonadiabatic couplings (i.e., the first derivative coupling), e.g., when it encounters a CI. This picture underlies our understanding of photochemical and photophysical processes.   
Modeling the conical intersection dynamics in this  Born-Huang framework requires propagating $N \ge 2$ nuclear wave packets  in $N$   adiabatic potential energy surfaces coupled by nonadiabatic coupling terms. The second derivative coupling is usually neglected, despite it is singular at CIs. Its importance to nonadiabatic excited state dynamics has  been emphasized \cite{meek2016}.
Although widely used in simulating excited state dynamics and ultrafast spectroscopy \cite{keefer2021a} , 
 the adiabatic  representation suffers from a major limitation that both the first and second derivative couplings, being proportional to the inverse of the energy gap,  diverge at CIs where energy gap vanish. This  makes the adiabatic representation inappropriate for wavepacket dynamics simulations \cite{guo2016}. 
To avoid such singularities, diabatic representation can be employed. While exact diabatization does not exist except for diatomics due to topological obstruction \cite{mead1982, tannor2007}, various quasi-diabatization methods based on different criteria   have been proposed \cite{baer2006, nakamura2002, subotnik2010, subotnik2008}. 
%The main idea is to transform nonadiabatic couplings into a
 One common approach is the adiabatic-to-diabatic transformation. Unlike adiabatic states, the diabatic states retain the electronic character and  are smooth with respect to the nuclear coordinates. The quasidiabatization does not transform all nonadiabatic couplings away, with the residual couplings often neglected. This approximation is not always valid \cite{choi2021}. 
Another widely used diabatic approach uses the crude adiabatic representation, which requires only electronic states at a single reference geometry, e.g., at the Franck-Condon point.  Vibronic coupling models developed under this representation is suitable for small amplitude motion such as internal conversion and singlet fission where the nuclear motion does not deviate far from the reference geometry in the course of photodynamics \cite{aleotti2021b}. Conventional  exact quantum dynamics methods using direct product grids or basis functions can be directly applied to such models with a few, typically less than ten, nuclear degrees of freedom \cite{farfan2019, farfan2020}. Recent developments in the time-dependent density-matrix renormalization group, or more generally, tensor network state-based methods, and multiconfigurational time-dependent Hartree allows modeling vibronic models with dozens of vibrational modes \cite{ma2022, meyer1990, beck2000}. However, the crude adiabatic representation cannot describe photochemical reactions which involves large amplitude motion of  molecular structures as in photoisomerization and photo-dissociation. 
 It is  desirable to develop a representation that directly employs adiabatic electronic states without  diabatization, yet free of the singularities due to CIs and can describe large amplitude structural changes. Previous attempts include the moving crude adiabatic representation that employs the moving Gaussian basis \cite{maskri2022, zhou2019}, where the adiabatic electronic states are defined at the center of each Gaussian basis.

We previously introduced a local diabatic representation (LDR) as an accurate and singularity-free framework for modeling CI wavepacket dynamics directly with adiabatic electronic states at specific nuclear geometries.  The reference nuclear geometries are  determined by a discrete variable representation of the nuclear position operators   \cite{gu2023b}. 
 The nonadiabatic transitions and geometric phase are all accounted for through the electronic overlap matrix (EOM), that is, the overlap matrix between adiabatic electronic states at different nuclear geometries.  This overlap matrix is bounded unlike derivative couplings, and it does not even require the adiabatic electronic states to be smooth with respect to the nuclear coordinates, which is a necessary condition to define the non-adiabatic couplings.
 The LDR has several advantages from conceptual and practical perspectives. 
 It avoids both first and second derivative couplings, and are thus free of the singularities they introduce at the CI. Moreover, since the LDR only requires the adiabatic states, it is straightforward to combine it with  well-established quantum chemistry methods including both Hartree-Fock-based and density function theory-based methods.

Here we first show that it is possible to directly apply the explicit split-operator algorithm to the full molecular propagator in the LDR. The split-operator method is restricted to Hamiltonians in which the kinetic and potential energy operators are separable \cite{choi2019, choi2020}.  Therefore,  in the conventional adiabatic representation, the split-operator method cannot be used  due to the presence of derivative couplings. 
%The  propagation is shown to be equivalent to the equation of motion from variational  principle owing to the fact that EOM is a projection operator. 
In comparison to the integrators based on finite difference approximation such as Euler, second-order difference, and Runge-Kutta, the split operator algorithm conserves the vibronic wavefunction norm and molecular energy exactly. Moreover, 
the time step allowed by the splitting operator method is one order of magnitude larger than our previous implementation with the fourth-order Runge-Kutta method. 
We then combine the LDR  and the uniform-grid discrete variable representation \cite{colbert1992} by employing the Fourier series as  the primitive nuclear basis set to  make the LDR method a universal solver for CI dynamics in arbitrary coordinates.  
%We previously used the coherent states (or, more generally, Gaussian wavepackets) as the primitive nuclear basis set.
Comparing to coherent states used in previous studies, the Fourier series does not suffer from linear dependence due to orthogonality and shows a monotonic convergence  increasing the number of grid points, whereas the nonorthogonal coherent states (or, more generally, Gaussian wavepackets) may encounter the linear dependence problem and requires a fine tuning regarding where to place the basis sets.  
%Moreover, while being a diabatic representation, it employs the adiabatic electronic eigenstates. 
%This representation removes the singularity because it does not involve nonadiabatic couplings.
The LDR method is implemented in a Python-based in-house code PyQED, relying heavily on the  Numpy module \cite{harris2020}. 
An illustration of the LDR is shown for a two-dimensional linear vibronic coupling model to compare with exact quantum dynamics results obtained by a conventional split-operator method in the diabatic representation.  
%and for the vibronic model of pyrazine molecule. 
It is found that the LDR results with adiabatic states is in excellent agreement with the exact results. 

This paper is structured as follows.  Sec. \ref{sec:theory} first presents the underlying theory of LDR with an emphasis of the properties satisfied by the EOM.  It is then shown that how the LDR enables the Strang splitting to be directly applied to the molecular propagator, leading to an efficient and numerically stable method for integrating the time-dependent molecular \se. 
\sec{sec:dvr} shows how to combine the LDR with the discrete variable representation with Fourier basis. 
\sec{sec:model} illustrates the utility of the new method for a two-dimensional vibrational coupling model.
\cref{sec:conclude} concludes.   

\section{Theory} \label{sec:theory}

\subsection{Local Diabatic Representation}

%We first choose a basis set $\set{\chi_\mu(\bf R)}$ to describe the nuclear motion.   We construct the  localized basis functions by diagonalizing the position matrix elements $x_{\mu \nu} = \braket{\chi_\mu | x|\chi_\nu}$. This is  discrete variable representation \cite{light2000}.
%% By construction, the basis set are the maximally localized states.
%%$\ket{x_n}$ are eigenstates of position operator, and hence localized in configuration space.
%The electronic Hamiltonian

%where $\chi_n(\bf R)$ is a nuclear basis centered at $\bf R_n$, that is,
%\be
%\braket{\chi_n | \bf R | \chi_n} = \bf R_n.
%\ee.
%$\phi_\alpha(\bf r, \bf R_n)$ is the $\alpha$th adiabatic electronic state $H_\text{BO}(\bf R_n)\phi_\alpha(\bf r; \bf R_n) = E_\alpha(\bf R_n) \phi(\bf r; \bf R_n) $ with $H_\text{BO}(\bf R) = H - \hat{T}_\text{n}$ the electronic Hamiltonian, the full molecular Hamiltonian subtracting the nuclear kinetic energy operator.
%Here $E_\alpha(\bf R)$ are the adiabatic potential energy surfaces.
%This ansatz can be understood as an expansion of the full molecular wavefunction in terms of the electron-vibrational (vibronic) basis set  $\set{\phi_\alpha(\bf r; \bf R_n) \chi_n(\bf R) }$ with expansion coefficients $C_{n\alpha}(t)$.
%
%It follows that
%\be
%H_{BO}(\bf r, \bf R) \ket{\bf R_\alpha} = H_{BO}(\bf r; \bf R_n) \ket{\bf R_\alpha}
%\label{eq:110}
%\ee

The LDR is constructed by firstly choosing a,  not necessarily orthogonal,  primitive nuclear basis set $\set{\zeta_\mu(\bf R)}$. For multi-dimensional systems, a straightforward choice is the direct product of one-dimensional basis functions, $\ket{\zeta^1_{\mu_1}} \otimes \ket{\zeta^2_{\mu_2}} \cdots$, where $\mu_I$ goes through all bases for $I$th nuclear degree of freedom.  This is followed by constructing the localized basis functions by diagonalizing the position matrix
\be
\bf X^I \ket{\chi_{n_I}^I} = R^I_{n_I} \ket{\chi_{n_I}^I}
\ee
with $ X^I_{\mu \nu} = \braket{\zeta^I_\mu | R_I |\zeta^I_\nu} = \int \dif R_I \zeta^{I*}_\mu(R_I) R_I \zeta^I_\nu(R_I) $ is the $I$th position matrix represented in the primitive basis set,  $\ket{\chi^I_{n_I}}$  the $n_I$th eigenstate with eigenvalue $R^I_{n_I}$.
Since all position operators commute with each other, they share common eigenstates. For product basis set, they are simply the direct product of the eigenstates of each position operator, i.e.,
\be
\ket{\chi_{\bf n} } %\equiv \ket{n_1, n_2, \dots, n_D} 
= \ket{\chi^1_{n_1}} \otimes \ket{\chi^2_{n_2}} \cdots  \otimes \ket{\chi^D_{n_D}}
\ee
where $\bf n = \set{n_1, n_2, \dots n_D}$ is multi-index, and $\ket{\chi^I_{n_I}}, I = 1, \dots, D$ is the $n_I$th eigenstate of position operator $R_I$, and localized in configuration space.

%This basis set defines a resolution of identity of the nuclear Hilbert space $I = \sum_{\mu, \nu} \ket{\chi_\mu} \del{S^{-1}}_{\mu \nu} \bra{\chi_\nu}$ where $S_{\mu \nu} = \braket{\chi_\mu |\chi_\nu}$ is the overlap matrix.

%It follows that
%\be
%H_{BO}(\bf r, \bf R) \ket{\bf R_\alpha} = H_{BO}(\bf r; \bf R_n) \ket{\bf R_n} \label{eq:111}
%\ee where $H_\text{BO}$ is the electronic Hamiltonian. \eq{eq:111} is valid under the finite basis representation.
In the LDR,  the ansatz for the full molecular wavefunction is  given by
\be
\Psi(\bf r, \bf R, t) = \sum_{\bf n} {\sum_\alpha C_{\bf n\alpha}(t) \phi_\alpha(\bf r; \bf R_{\bf n})} \chi_{\bf n}(\bf R) \equiv \sum_{\bf n \alpha} C_{\bf n \alpha}(t)\ket{\bf n \alpha}
\label{eq:100}
\ee
where $\phi_\alpha(\bf R_{\bf n})$ is  the $\alpha$th adiabatic electronic eigenstate of the electronic Born-Oppenheimer Hamiltonian, the full molecular Hamiltonian excluding the nuclear kinetic energy $H_{\text{BO}}(\bf R) = H - \hat{T}_\text{N}$, at the nuclear geometry $\bf R_{\bf n} = \del{R^1_{n_1}, R^2_{n_2}, \cdots, R^D_{n_D}}$ with energy $E_\alpha(\bf R_{\bf n})$, i.e.,
\be 
H_\text{BO}(\bf R_{\bf n})\phi_\alpha(\bf r; \bf R_{\bf n}) = E_\alpha(\bf R_{\bf n}) \phi_\alpha(\bf r; \bf R_{\bf n})
\ee
As the vibronic basis set are orthonormal, it defines a resolution of the identity for the vibronic Hilbert space 
\be 
\mc{I} = \sum_{\bf m \beta} \proj{\bf m \beta}.
\label{eq:roi}
\ee 
%, where $\alpha$ labels the electronic states.

%This is a discretized form of the gauge transformation in the Born-Huang expansion, $U = e^{i \theta(\bf R)}$.
%the $\theta_n$ can be independently chosen arbitrarily because, as shown below, our dynamical equation does not involve nuclear derivative of the electronic wavefunctions.

%if we use an ansatz
%\be
%\Psi(\bf r, \bf R, t) = \sum_n \del{\sum_\alpha C_{n\alpha}(t) \phi_\alpha(\bf r; \bf R_n)} \chi_n(\bf R; \bf R_n, \bf P_n)
%\ee
%where $\chi$ is a Gaussian wavepacket centered at $\bf R_n$.
%This avoids the singular NAC at CIs.

%\be H \ket{\phi} = E \ket{\phi}
%\ee
%
%Using a set of GWP as basis,
%\be
%\ket{\phi_n} = \sum_\alpha \ket{\alpha} C_{\alpha n}
%\ee
%\be
%H_{\beta \alpha} C_{\alpha n} = E_n S_{\beta \alpha} C_{\alpha n}
%\ee
%
%For the potential energy operator
%\be
%x_{\beta \alpha} = \braket{\beta | x|\alpha}
%\ee
%%if we diagonalize $x$ by
%%\be
%%x = U^\dag \Lambda U
%%\ee
%%\textbf{This is wrong!}
%We should do
%\be
%x \ket{x_n} = x_n \ket{x_n}
%\ee
%say
%\be
%\ket{x_n} = C_{\alpha n} \ket{\alpha}
%\ee
%then
%\be
%x C =  S C X
%\ee

%The local diabatic picture solves the singularity problem, how to address the exponential scaling?

%our ansatz reads
%\be
%\ket{\Psi} = \sum_\alpha \del{\sum_n C_{n\alpha} \ket{\psi_n(\bf R_\alpha)} } \ket{\bf R_\alpha}
%\label{eq:101}
%\ee
%where $\Psi$ is the molecular state, $\hat{R}_\mu \ket{R_\alpha} = R_\mu \ket{\bf R^\alpha}$

Inserting \eq{eq:100} into the molecular \tdse\ $i \pd{\Psi}{t} =H\Psi$ with 
the molecular  Hamiltonian 
$
H = \hat{T}_\text{N} + H_\text{BO}(\bf R),
$
 and left multiply $\bra{\bf m \beta}$ yields the equation of motion for the expansion coefficients

\be
i \dot{C}_{\bf m \beta}(t)
%\phi_\alpha(\bf r; \bf R_n) \chi_n(\bf R; \bf R_n, \bf P_n)
= E_\beta(\bf R_{\bf m})C_{\bf m \beta}(t)+  \sum_{\bf n, \alpha} T_{\bf m \bf n}A_{\bf m \beta, \bf n \alpha} C_{\bf n\alpha}
\label{eq:main}.
\ee
Here
\be 
T_{\bf m \bf n} = \braket{\chi_{\bf m}| \hat{T}_\text{N} | \chi_{\bf n}}_{\bf R} = \sum_I - \frac{1}{2M_I} \Braket{\chi^I_{m_I} | \grad_I^2 | \chi^I_{n_I}} \otimes \bigotimes_{J \ne I} \delta_{m_J, n_J}
\ee 
is the kinetic energy operator matrix elements and the electronic overlap matrix (EOM)
\be
A_{\bf m\beta, \bf n \alpha} = \braket{\phi_{\beta}(\bf R_{\bf m}) | \phi_{\alpha}(\bf R_{\bf n}) }_{\bf r},
\ee
where $\braket{\cdots}_{\bf r}$ ($\braket{\cdots}_{\bf R}$) denotes the integration over electronic (nuclear) degrees of freedom.
In deriving \eq{eq:main}, we have made use of 
\be H_\text{BO}(\bf R) \ket{\bf n \alpha} \approx E_\alpha(\bf R_{\bf n}) \ket{\bf n \alpha}  
\ee 
as the nuclear state $\ket{\bf n}$ is an eigenstate of all position operators.

%From a practical perspective, \eq{eq:main} offers several advantages.
In the adiabatic representation, both derivative couplings \be \bf F_{\beta \alpha} = \braket{\phi_\beta(\bf R)|\bm \grad | \phi_\alpha(\bf R)}, \bf G_{\beta \alpha} = \braket{\phi_\beta(\bf R)|\bm \grad^2 | \phi_\alpha(\bf R)}
\ee 
 diverges at the CI point.
In the LDR, although we have employed the adiabatic electronic states, \eq{eq:main} does not contain any singularities because the nuclear kinetic energy operator does not operate on the electronic states. Therefore, in contrast to wavepacket dynamics in the adiabatic representation, the singularity of the nonadiabatic coupling at the CI  will not affect the simulations. In fact, the CI configuration can be safely  included in the basis set. % which does not affect the simulations.
%This is in contrast to the adiabatic representation where the CI cannot be included.
In practice, the kinetic energy matrix elements can be analytically calculated within the primitive basis set and then transformed into the position eigenstates. The adiabatic electronic states can be computed with well-established electronic structure methods.

\Eq{eq:main} can be applied for an arbitrary gauge. It does not require the electronic wavefunction to be continuous with respect to the nuclear geometry. This means that electronic structure calculations at different geometries can be directly used without further smoothing procedure.

\subsection{Electronic overlap matrix}
The overlap matrix $A$ encodes \emph{all} nonadiabatic effects including geometric phase. The adiabatic approximation in the LDR amounts to 
\be
A_{\bf m \beta, \bf n \alpha } \approx \delta_{\beta \alpha} \delta_{\bf m \bf n}  
\ee 
so that there are no nonadiabatic transitions and leads to the adiabatic Born-Oppenheimer dynamics.
Below we discuss some of its properties. 
\begin{enumerate}
	\item  Hermiticity - 
%It admits some interesting properties. 
By definition, it is  Hermitian
\be
A_{\bf m \beta, \bf n \alpha } = A_{\bf n \alpha, \bf m\beta}^*.
\ee
This holds even  when the electronic wavefunction is complex. For any $\bf n$,
$
A_{\bf n \beta, \bf n \alpha} = \delta_{\beta \alpha}
$
due to orthonormality of the adiabatic electronic states. 

\item Gauge transformation - 
There is a U(1) local phase gauge freedom in the LDR ansatz [\eq{eq:100}],
\be
\ket{\chi_{\bf n}'} = e^{i \theta_{\bf n}} \ket{\chi_{\bf n}}, ~~~ \ket{\phi_\alpha'(\bf R_{\bf n})} = e^{-i\theta_{\bf n}} \ket{\phi_\alpha(\bf R_{\bf n})}
\label{eq:gauge}.
\ee
Under this gauge transformation in \eq{eq:gauge}, 
\be 
A'_{\bf m \beta, \bf n \alpha} =  A_{\bf m \beta, \bf n\alpha} e^{-i \del{\theta_{\bf n} - \theta_{\bf m}}}.
\ee Thus, it is not gauge-invariant.    The  Wilson loop  $W_\alpha = \prod_\gamma A_{i \alpha, i + 1\alpha}$ for a loop $\gamma$ encircling a CI, reflects the geometric phase and is  gauge invariant.

\item Projection operator - 
Another important property of the overlap matrix is that
\be
\bf A^2 = \bf A
\ee
This means that the EOM is a projection operator, and the eigenvalues are $\pm 1$. 

\item Connection to  nonadiabatic coupling - 
The EOM is closely related to the nonadiabatic coupling, which can be written as 
\be
F^I_{\beta \alpha}(\bf R) %= \braket{\phi_\beta(\bf R)| \bm \grad| \phi_\alpha(\bf R)} 
= \lim_{\delta \rightarrow 0} \frac{1}{\delta} \del{\phi_\beta(\bf R) | \phi_\alpha(\bf R + \bf e_I \delta)   - \bf I}
\ee  
where the adiabatic electronic states are defined in a smooth gauge and $\bf e_I$ is a unit vector in the $I$th direction.  In a continuum limit, the overlap matrix is generalized to be  
\be
A_{\beta \alpha}(\bf R, \bf R')  \equiv \braket{\phi_\beta(\bf R)|\phi_\alpha(\bf R')}
\ee 

Choosing a path $\gamma$ that starts from $\bf R$ and ends at $\bf R' $,  discretizing into $N$ segments, and inserting a resolution of identity [\eq{eq:roi}] into each node leads to 
 \be 
 \bf A(\bf R, \bf R')  =  \lim_{N\rightarrow \infty} \prod_{i=0}^{N-1} \braket{ \phi_{\alpha_i}(\bf R_i) | \phi_{\alpha_{i+1}}(\bf R_{i+1})} = \prod_{i=0}^{N} e^{\sum_I { \Delta_{i}^I \bf F_I(\bf R_i)}} = \mc{P} e^{\int_\gamma \bf F(\bf R) \cdot  \dif \bf R} 
 \ee 
 where $ \bf R_0 = \bf R, \bf R_{N} = \bf R', \bm \Delta_i = \bf R_{i+1} - \bf R_i$, $\mc{P}$ is the path-ordering operator. Note that the EOM is path-dependent. If we choose two paths $\gamma_1$ and $\gamma_2$ with the same end points, the path-independence requires that the loop integral of non-adiabatic coupling along $\gamma_{12} = \gamma_1 - \gamma_2$, tracing path $\gamma_1$ first and then tracing path $\gamma_2$ backwards in time,  has to vanish. This is in contradiction to  geometric phase.  This implies that the EOM function  cannot be made globally continuous in the presence of CIs. 
 
% From \eq{eq:111}, we realize that 
% \be 
% e^{\bf F^\mu(\bf R) \Delta} = \braket{\phi_\beta(\bf R) | \phi_\alpha(\bf R + \Delta \bf e_\mu)}
% \ee 

\end{enumerate}
%Any nuclear basis set can be used. Depending on the nuclear degree of freedom, different basis may be more convenient and require less bases to converge than the other. For example, a Fourier basis may be convenient for rotation %angular coordinates
%whereas local Gaussian bases including  coherent states may be useful for describing vibrational dynamics.
%In the limit of a single Gaussian basis, the position eigenvalue is simply the center, our ansatz reduces to the

%\section{Multidimensions}

%say we have $N$ nuclear dofs, then the position vector is $\bf R$, the problem when we diagonalize $R_j$, are we getting the same eigenstates? in principle, we should caz $[R_i, R_j] = 0$. but in the FBR, does it hold?
%

\subsection{Strang spiting of the total molecular propagator} 
 The split-operator method with fast Fourier transform is an efficient method for propagating a nuclear wavepacket in a single potential energy surface. It can also be used for the  wavepacket dynamis of a vibronic coupling model Hamiltonian in the diabatic representation. However,
in the adiabatic representation, the split-operator method cannot be used due to the presence of the non-adiabatic coupling terms that involves both position and momentum operators \cite{choi2019}.

We show that  the split-operator method can be applied to the LDR ansatz. While we previously solved \eq{eq:main} by the Runge-Kutta method, we found that it is advantages to apply directly the molecular propagator $e^{- i H t}$ to the ansatz in \eq{eq:100}

Applying the second-order Strang splitting for the total molecular propagator
\be
e^{-i H \Delta t} \approx  e^{- i H_\text{BO}(\bf R) \Delta t/2}   e^{- i \hat{T}_\text{N} \Delta t/2} e^{- i H_\text{BO}(\bf R) \Delta t/2} + \mc{O}(\Delta t^3)
\label{eq:spo}
\ee
The symmetric splitting reduces the error to the third-order \cite{strang1968}. 

The first exponential operator due to  $H_\text{BO}(\bf R)$ can be applied to the nuclear wavefunction
\be
\sum_{\bf n, \alpha} e^{-i H_\text{BO}(\bf R) \Delta t/2} C_{\bf n\alpha}(t) \ket{\bf n \alpha}  \approx \sum_{\bf n, \alpha} C_{\bf n\alpha}(t) e^{-i E_{\alpha}(\bf R_{\bf n}) \Delta t/2 } \ket{\bf n\alpha}
\ee
where we have made use of the fact that $\ket{\bf n}$ is an eigenstate of the position operators.
For the nuclear kinetic energy operator, since $\hat{T}_\text{N}$ only applies to the nuclear basis functions,
\be
 e^{- i  \hat{T}_\text{N} \Delta t} C_{\bf n \alpha}\ket{\bf n \alpha } = \sum_{\bf m, \beta } \sum_{\bf n, \alpha} \ket{\bf m \beta}\braket{ \bf m\beta| e^{- i T\Delta t} | \bf n \alpha }   = % \braket{\phi_\beta(\bf R_m) | \phi_\alpha(\bf R_n)}_{\bf r}
 \sum_{\bf n, \alpha}  A_{\bf m \bf n}^{\beta \alpha} \braket{\bf m | e^{-i T \Delta t}| \bf n}_{\bf R}  C_{\bf n\alpha} %= A_{n_{i+1} \alpha_{i+1},, n_i \alpha} \braket{\chi_m | e^{-i T_\text{N} \Delta t}| \chi_n}_{\bf R}
\ee
where in the second step we have inserted  resolution of the identity.
The split-operator method in \cref{eq:spo}  has the  advantages of norm- and energy-conservation in comparison to the Runge-Kutta method.

%\subsubsection{Comparison to }

%At first glance, \eq{eq:spo} seems to be inconsistent with \eq{eq:main} obtained from variational principle. If we apply  the split-operator method to \eq{eq:main}, it yields 
%\be
% %C_{\bf m \beta}(t + \Delta t) =  e^{-i E_{\bf m \beta} \Delta t/2}  e^{-i A_{\bf m \beta, \bf n \alpha} T_{\bf m \bf n}\Delta t} e^{-i E_{\bf n \alpha} \Delta t/2} C_{\bf n \alpha}(t)
%\bf C(t + \Delta t)  = e^{-i \bf V \Delta  t/2} e^{-i \bf \mc{T} \Delta t}  e^{-i \bf V \Delta t /2}\bf C(t)
%\label{eq:111}
%\ee
%where $\mc{T}_{\bf m\beta, \bf n \alpha} = T_{\bf m \bf n} A_{\bf m\beta, \bf n \alpha}, V_{\bf m \beta, \bf n \alpha} = \delta_{\beta \alpha}  \delta_{ \bf m \bf n}E_\alpha(\bf R_{\bf n})$. The equivalence between \cref{eq:111} and \cref{eq:spo} can be shown by first realizing that 
%\be
%%[A, \mc{T}] = 0 
%\mc{T}^2 = A_{\bf l \gamma, \bf m \beta} T_{\bf l \bf m} A_{\bf m \beta, \bf n \alpha}  
%\ee 

\subsection{Fourier basis} \label{sec:dvr}
The Fourier basis is widely used in  adiabatic wave packet dynamics on a single potential energy surface \cite{colbert1992}.
For vanishing boundary conditions $\xi(a) = \xi(b) = 0$ in an arbitrary range $x \in [a, b]$, the Fourier bases (or particle in-a-box eigenstates) become  \cite{beck2000}
\be
\xi_n(x) = \sqrt{\frac{2}{L}} \sin\del{\frac{ n \pi (x-a)}{L}},  n = 1,  2 \cdots,  N-1
\ee
where $L = b - a$.  
The multidimensional Fourier basis can be constructed by tensor product.
 Instead of computing the position matrix elements, we compute the operator  $z = \cos(\pi (x-a)/L)$. The matrix elements can be calculated to be 
\be
 \braket{\xi_m| z |\xi_n } = \half \del{\delta_{m+1, n} + \delta_{m-1,n}} 
.\ee
This tridiagonal matrix is often encountered in the tight-binding model for condensed matter systems.
%$e^{i 2\pi (x-a)/L}$.  eigenstates of the $X_{nm} = $ reads
The eigenvalues read
\be
z_k = \cos(k\pi/(N+1))  , k =1, 2, \cdots, N
\ee 
The corresponding grid points   are then  
\be
%x_n = n \frac{2\pi}{2N+1}, n = 0, \cdots, 2N
x_n = a + n\Delta x, ~~~ n = 1, 2, \dots, N
\ee
where $\Delta x = L/(N+1)$. 
As the Fourier basis itself are orthonormal, the position eigenstates are approximately \cite{colbert1992}
\be
\chi_n(x) %= \sum_n \xi_{n}(x) \xi_n(x_n)
 = \sqrt{\frac{1}{\Delta x}} \text{sinc} \del{ \frac{\pi \del{x - n\Delta x} }{\Delta x} }
\ee
In  Fourier basis, the kinetic energy matrix  can be calculated analytically, which for $a \rightarrow -\infty, b \rightarrow \infty$ is approximately  \cite{colbert1992}
\be
T_{nn'} =% \equiv \braket{\chi_n | -\frac{1}{2m} \grad^2 |\chi_{n'}}=  
\frac{1}{2m \Delta x^2} \del{-1}^{n-n'} 
\begin{cases}
%	\frac{N(N+1)}{3} & n = n' \\
\pi^2/3 & n = n' \\
%	\frac{\cos(\pi (n-n') / (2N+1))}{2 \sin^2(\pi (n-n')/(2N+1))} & n \ne n'
\frac{2}{(n-n')^2}  & n \ne n'
\end{cases}
\ee
For angular coordinate $\theta \in [0, 2\pi]$ requiring  
%%It is straightforward to applying the following method to any range by translation and scaling.    
 periodic boundary condition $\xi(0) = \xi(2\pi)$, we can employ Fourier series
\be
\xi_n(\theta) =  \sqrt{\frac{1}{2\pi}} e^{i \theta n},  n = 0, \pm 1,  \pm 2 \cdots,  \pm N
\ee

%The map from a natural index to Fourier components is
%\be
%\sigma(n) = \begin{cases}
	%	-n/2 & n \in \text{even} \\
	%	(n+1)/2 & n \text{ is odd}
	%\end{cases}
	%\ee
%	Transformation from the Fourier basis to the DVR set is given by the discrete Fourier transform.

%	The momentum matrix elements can be easily computed
%	as
%	\be
%	\braket{\chi_n | p |\chi_m} = n \delta_{nm}
%	\ee
%
%	The KEO for the rotational coordinates is usually more complicated than the cartesian coordinates.

\section{Model and Computational details} \label{sec:model}

\begin{figure}[hbpt]
	\includegraphics[width=0.8\textwidth]{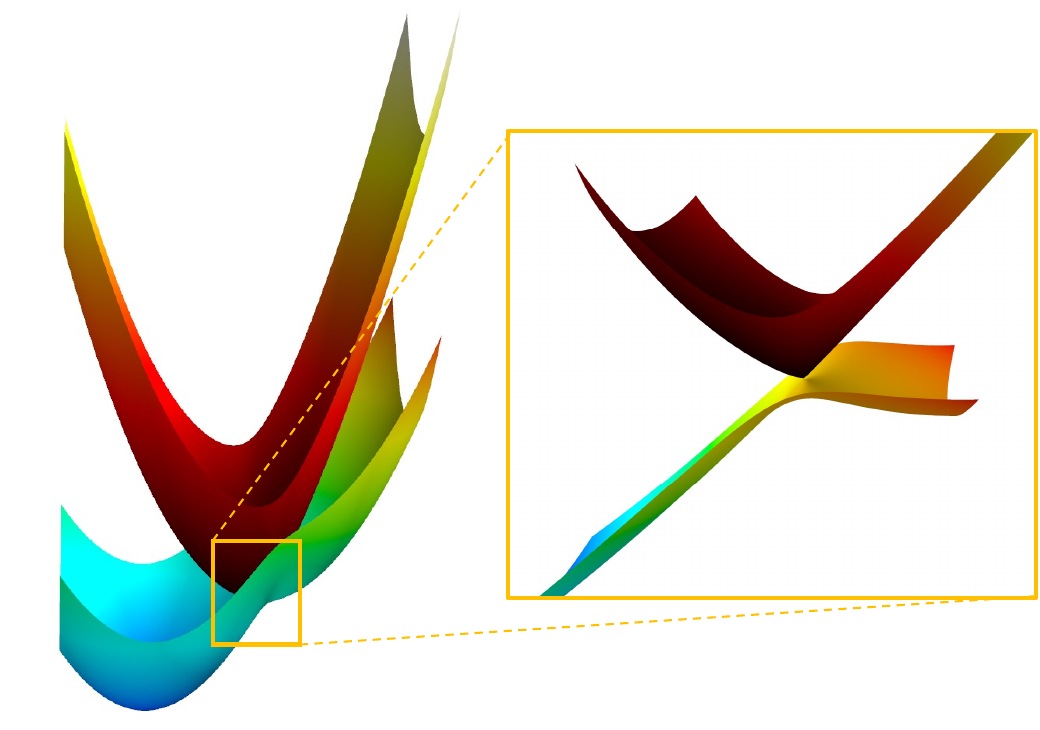}
\caption{Adiabatic potential energy surfaces of the vibronic model. Right is a zoom in of the conical intersection. Here $E_1 = 2, \lambda = 0.2$.}
\label{fig:apes}
\end{figure}

We demonstrate the utility of the LDR/SPO/Fourier (split operator method with Fourier basis) for nonadiabatic conical intersection wavepacket dynamics by a vibronic  coupling Hamiltonian, which reads
\be
H = \frac{p_x^2   + p_y^2  }{2} + % \frac{ (x+1)^2 +y^2) }{2} \proj{0}  +  V_1(\bf R)\proj{1} + \lambda y \del{\ket{0}\bra{1} + \ket{1}\bra{0}}
\sbr{\begin{array}{cc}
	\half \del{(x+1)^2 +y^2)} & \lambda y \\ 
	\lambda y & \half\del{(x-1)^2 +y^2}  +E_1
\end{array}}
\ee
where $x$ is the tuning mode and $y$ is the coupling mode, $p_x$ and $p_y$ are respectively the momentum for $x$ and  $y$, $\lambda$ is the inter-state coupling constant, $E_1$ is the minimum energy difference between two diabatic surfaces. % transition energy at $(0, 0)$.
 The APESs, obtained by diagonalizing the diabatic potential energy surfaces, are depicted in \fig{fig:apes}. It exhibits a CI at $ (-E_1/2, 0)$.

The information of molecular structures are contained in the nuclear reduced density matrix %reads
\be
\rho^\text{N}(t) = \sum_{n,m} \sum_{\beta, \alpha} C_{m\beta}^* A_{m\beta, n\alpha} C_{n\alpha} \ket{\bf n}\bra{\bf m} 
\ee
The uniform grids are particularly convenient to compute the  nuclear density. For the total nuclear probability density 
\be
\begin{aligned}
\rho(\bf R_{\bf l}, t) &= \braket{\bf R_{\bf l}|\rho^\text{N}(t)|\bf R_{\bf l}} = \sum_{\bf n, \bf m} \sum_{\beta, \alpha} C_{\bf m\beta}^* A_{\bf m\beta, \bf n \alpha} C_{\bf n \alpha} \chi_{\bf n}(\bf R_{\bf l}) \chi^*_{\bf m}(\bf R_{\bf l}) \\ 
& = \sum_{\beta, \alpha} C_{\bf l\beta}^* A_{\bf l \beta, \bf l \alpha} C_{\bf l \alpha} \\
&= \sum_{\alpha} \abs{C_{\bf l \alpha}}^2 \\ 
\end{aligned}
\label{eq:111}
\ee
where we have used the property of the sinc functions $\chi_{\bf n}(\bf R_{\bf m}) = \delta_{\bf n \bf m}$.
It is clear from \eq{eq:111} that the total nuclear probability density is the sum of the nuclear probability density on each APES.  
The electronic observables can be computed from the electronic reduced density matrix, which is given by 
\be
\rho^\text{e}(t) = \sum_{\alpha, \beta} \sum_{\bf n} C_{\bf n \alpha}(t) C_{\bf n \beta}^*(t) \ket{\phi_\alpha(\bf R_{\bf n})} \bra{\phi_\beta(\bf R_{\bf n})}. 
\ee

The reference exact nonadiabatic quantum wavepacket dynamics is performed with the conventional split-operator method and fast Fourier transform on a uniform grid  in the diabatic representation \cite{kosloff1988}, i.e., 
\be
\chi^{(\text{d})}(t+\Delta t) = e^{-i \bf V \Delta t/2} e^{-i \hat{T}_\text{N} \Delta t} e^{-i \bf V \Delta t/2} \chi^{(\text{d})}(t)
, \ee
where $\bf V$ is the diabatic surfaces  and the kinetic energy propagator is performed by a fast Fourier transform.   
The time step is $\Delta t = 0.5$ a.u. The nuclear wave packets in the adiabatic representation is computed via the diabatic-to-adiabatic transformation matrix
\be
\chi^{(\text{a})}_{\bf n \alpha}(t) =  \sum_\beta U_{\bf n}^{\alpha \beta} \chi^{(\text{d})}_{\bf n \beta}(t)
\ee
where $U_{\bf n}$ is the diabatic-to-adiabatic unitary transformation matrix at nuclear configuration $\bf R_{\bf n}$. The transformation matrix  is obtained by diagonalizing the diabatic potential energy matrix for each configuration.  A uniform grid of 33 points in the range $[-6, 6]$ a.u. are used for both modes.  

\begin{figure}[hbpt]
	\includegraphics[width=0.6\textwidth]{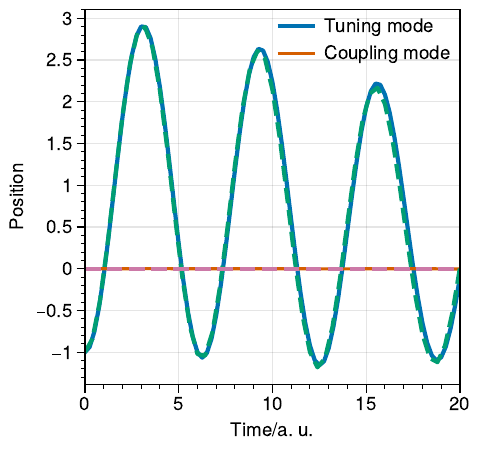}
	\caption{Expected positions of the tuning and coupling modes. Solid lines are LDR results while the dashed lines are exact results.}
\label{fig:x}
\end{figure}

\begin{figure}[hbpt]
	\includegraphics[width=0.6\textwidth]{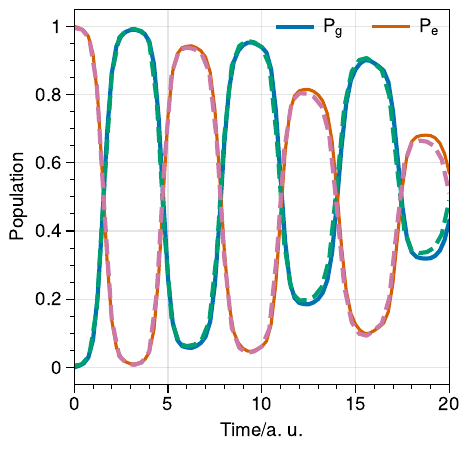}
	\caption{Adiabatic population dynamics exhibiting rapid population transfer between two adiabatic surfaces mediated by conical intersections. The molecule is initially in the excited state.}
	\label{fig:popu}
\end{figure}

\begin{figure}[hbpt]
	\includegraphics[width=1\textwidth]{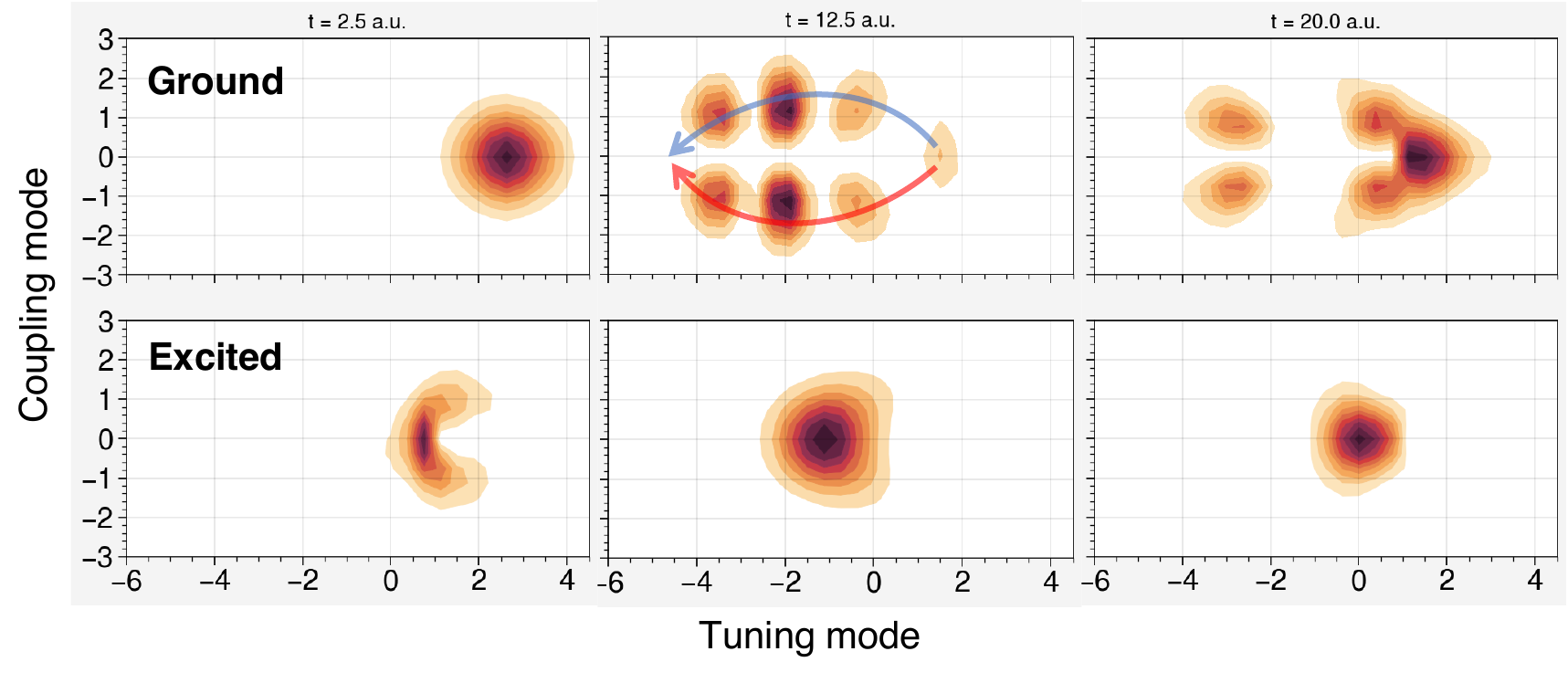}
	\caption{Snapshots of the nuclear wave packet dynamics in the adiabatic  electronic states at indicated times. The nuclear wavepacket in the ground potential energy surface shows clearly the geometric phase effect.}
	\label{fig:wp}
\end{figure}

The molecule  is initially vertically excited to state $\ket{1}$,  $\Psi_0 = {\pi}^{-1/2}  e^{- \half (x+1)^2 - \half y^2} \proj{1} $.
\cref{fig:popu} shows the adiabatic population dynamics for around four periods of the vibrational modes and \cref{fig:x} shows the expectation value of the tuning and coupling modes. Both are in exact agreement with the results obtained in the diabatic presentation.  As the nuclear wavepacket traverses through the CI, there is a rapid population transfer between the two adiabatic states. \cref{fig:wp} depicted three snapshots of the the nuclear wavepackets on the adiabatic electronic states at indicated times. The nuclear wavepacket in the ground state shows a nodal line along $y = 0$,  a hallmark of the geometric phase effect. 

The time step used in the current simulations $\Delta t = 0.5$ a.u. is  one order of magnitude larger than the one allowed by our previous implementation using the fourth-order Runge-Kutta method $\Delta t = 0.02$ a.u.. This is a significant advantage provided by the split-operator algorithm.

%\subsection{Lack of electronic coherence in LVC models}
%
%The electronic coherence is zero because the nuclear wavepacket in the ground state is odd under $x \rightarrow -x$ while the one in the excited state is even.  

\section{Conclusion} \label{sec:conclude}

 We have shown that the local diabatic representation (LDR) allows a straightforward applicaton of the  Strang splitting to the molecular propagator, which leads to a better integration algorithm for the correlated electron-nuclear quantum dynamics around conical intersections. We further showed how to combine the LDR with the uniform-grid discrete variable representation employing Fourier series.  
 The LDR can be interfaced straightforwardly with electronic structure packages, of which many can directly output the electronic wavefunction overlap.

The main limitation of the LDR method is the exponential scaling of computational cost with system size because it uses a direct product basis set. This not only increases the size of the Hamiltonian matrix but also requires a prohibitively large amount of electronic structure computations for ab initio modeling.  
This challenge cannot  be tackled  alone by simplifying the expansion coefficients $C_{\bf n \alpha}$  by e.g. a Hartree product or matrix product state because this does not reduce the number of grid points in the configuration space. Even for a Hartree-product, the number of electronic structure computations still scales exponentially with system size. The key to extend  LDR to high-dimensional ab initio conical intersection dynamics is to reduce the number of grid points. A promising strategy to do so is to use sparse grids which significantly reduces  the grid points, and hence the Hamiltonian matrix size \cite{hackbusch1985, avila2015, ahai2022}.  
This direction is currently under progress. 

\bibliography{../../cavity,../../optics,../../qchem,../../dynamics}

\end{document}